# Synaptic Learning and Memory Functions Achieved in Self-rectifying BFO Memristor under Extreme Environmental Temperature

Nan Yang [a], Zhong-Qi Ren [a], Zhao Guan [a], Bo-Bo Tian [a], Ni Zhong [a,b]*, Ping-Hua Xiang [a,b]*, Chun-Gang Duan [a,b] and Jun-Hao Chu [a,b]

[a] Key Laboratory of Polar Materials and Devices, Ministry of Education, Department of Optoelectronics, East China Normal University, Shanghai, 200241, China

[b] Collaborative Innovation Center of Extreme Optics, Shanxi University, Taiyuan, Shanxi 030006, China

E-mail: nzhong@ee.ecnu.edu.cn, phxiang@ee.ecnu.edu.cn

**Abstract**

Memristors have been intensively studied in recent years as promising building blocks for next-generation non-volatile memory, artificial neural networks and brain-inspired computing systems. Even though the environment adaptability of memristor has been required in many application fields, it has been rarely reported due to the underlying mechanism could become invalid especially at an elevated temperature. Here, we focus on achieving synaptic learning and memory functions in $BiFeO_3$ memristor in a wide range of temperature. We have proved the ferroelectricity of BFO films at a record-high temperature of 500 ℃ by piezoresponse force microscopy (PFM) measurement. Due to the robust ferroelectricity of BFO thin film, an analog-like resistance switching behavior has been clearly found in a wide range of temperature, which is attributed to the reversal of ferroelectric polarization. Various synaptic functions including long-term potentiation (*LTP*), depression (*LTD*), consecutive potentiation/depression (*P/D*) and spike-timing dependent plasticity (*STDP*) have been realized from -170 to 300 ℃, illustrating their potential for electronic applications even under extreme environmental temperature.

**Introduction**

Brain-inspired computation, also referred to as neuromorphic computing, which is the chief approach to engineer new computational architectures with low energy consumption and greater efficiency, has long attracted significant interest.[1-4] Efficiency and computational power of neuromorphic systems depends on the accurate implementation of the brain's building blocks, neurons and synapses.[5] The synapses act as an essential part of connecting the presynaptic and postsynaptic neurons in the brain nervous system, which play a key role in learning and memory (**Figure 1a**).[6-8] A compact electronic device which mimics biological synapse is the critical step to build massively artificial neural networks (i.e., machine/deep learning) for brain-inspired computation.[9-11] However, artificial neural networks are typically constructed using electronic components called complementary metal–oxide–semiconductor (CMOS) devices, which need to be intensively programmed using software. This results in formidable computing complexity and high power consumption. These drawbacks severely hinder the practical application of CMOS-based neural-network systems for power-hungry computing tasks.[11] To tackle these challenges, electronic components known as memristors have become promising candidates as a basic block for building artificial neural networks, which are highly parallel, fault-tolerance, energy-efficient, and event-driven information processing system.[12-15] The aim is to realize energy efficient artificial intelligence processes at the hardware level. Many research works have been made in building hardware implementation that incorporate Mott memristors[16-18], phase-change memristors[19-21] and redox memristors[12, 22-24] to emulate synapses by utilizing their tunable resistance (conductance) as synaptic weights. Device switching performance and reliability have been improved in optimal conditions. However, no reliable synaptic characteristics observed in memristors at a wide temperature range limits their potential application in harsh environments such as space exploration, deep drill, automotive industry, *etc*. Memristors based on ion moving or ionic valence changing[25-27] always exhibit a failure operation at high temperatures above 200 ℃ due to thermal instability.[28] Several groups also reported a remarkable degraded performance in the phase-change memristor under high temperature.[29-31] Moreover, the synaptic functions are expected to be ineffective at extremely low temperature since the threshold voltage exhibit a significant dependence on the operation temperature.[32, 33]

Recently, ferroelectric memristor as next-generation nonvolatile memristor (NVM) has attracted intensive attention, which have shown high resistive ON/OFF ratio, fast switching speed, satisfactory endurance and low energy consumption.[34-37] Ferroelectric materials exhibit two distinctive states of their spontaneous polarization, i.e. ferroelectric up (+$P$) and down (-$P$). These two states are defined by thermodynamics, that is, both states ideally correspond to free-energy-minimizing configuration.[38, 39] By changing the direction of applied electric field, i.e. polarity alternation, ferroelectric polarization can be reversed. In a metal-ferroelectric-metal (MFM) device, the ferroelectric region includes multi-domain and its remnant ferroelectric polarization is determined by the average polarization value. By controlling domain configuration, 'various polarization states' lying between +*P* and -*P* can basically be achieved. In this case, remnant polarization works as synaptic weight. In MFM device, we can switch the ferroelectric polarization to modulate the resistance.[34, 40, 41] Importantly, as a key aspect of ferroelectric memristor, ferroelectric polarization dependent resistance modulation could be observed when spontaneous polarization remains stable. Therefore, good environment/temperature adaptability can be expected in ferroelectric memristors if the ferroelectric phase could be stable in a wide range of temperature.

In this article, we report the ferroelectric memristor based on $BiFeO_3$ (BFO) films, which exhibits large electric polarization and high ferroelectric Curie temperature of ~1103 K.[42] Piezoresponse force microscopy (PFM) measurement exhibits a stable ferroelectricity of epitaxial BFO thin films up to a record-high temperature of 500 ℃. A significant resistance switch (RS) characterization was identified over a wide temperature range. It was clearly demonstrated that the memristor resistance (conductance) was modulated by the ferroelectric reversal of BFO films. The BFO-based memristors successfully mimic the key features of biological synapses, with long-term potentiation (*LTP*), depression (*LTD*), consecutive potentiation/depression (*P/D*) and spike-timing dependent plasticity (*STDP*) learning rule in a wide range of temperature from -170 to 300 ℃, which is expected to be extended to 500 ℃ due to the stable ferroelectricity observed at extremely high temperature.

**Mechanism of Device Operation**

Our ferroelectric memristor is based on metal-ferroelectric-metal (MFM) device, in which BFO film is sandwiched between SRO bottom and Au top electrodes. During all the electrical measurement, the voltages were applied on the Au top electrode and the SRO electrode was always grounded. In biology, synaptic weight may be changed from a presynaptic axon terminal to a postsynaptic dendrite terminal through electrical or electrochemical signals.[43, 44] BFO memristor is similar to biological synapse, and the synaptic functions are realized by changing electric stimuli. Figure 1a schematically illustrates the working principle of BFO memristor and the analogy to a biological neural system with typical synaptic structures. Figure 1b shows typical dc *I-V* curves, which exhibit a remarkable RS behavior with a diode-like rectification characteristic. It should be noted that RS could be observed when the amplitude of applied voltage is higher than a threshold value. Moreover, the high resistance state (HRS) remains constant and the low resistance state (LRS) depends on the voltage amplitude on the negative voltage sweeps, demonstrating the existence of multi-level resistance states in our BFO memristor.

To gain insight into the mechanism accounted for the RS behavior, we investigated the relation between the resistance and ferroelectric polarization at room temperature (RT). BFO memristor is characterized by applying a writing voltage pulses ($V_{write}$) to modify polarization and the memristor resistance was subsequently measured at a low pulse voltage ($V_{read}$ = -1.5 V, $t_{pulse}$ = 0.5 ms). Figure 2b shows a typical hysteresis curve of resistance vs $V_{write}$ (*R-V* loop), and its asymmetry should be noted. It is found that applying positive voltage pulses with increasing amplitude virtually does not change the resistance state of the device until a threshold voltage of $V_{th}^+$ = +2.5 V. As $V_{write}$ further increases beyond this threshold value, the resistance rises significantly by about 3 orders of magnitude (OFF state). When $V_{write}$ is then swept from +7 V to -7 V, the resistance is stable up to a negative threshold voltage $V_{th}^-$ = -3.5 V, and decreases with voltage amplitude when $V_{th}^-$ < -3.5 V. The *P-V* loop measurement was carried out at -170 °C, and a good ferroelectricity with asymmetric coercive voltages of about $V_{th}^-$ = -4 V and $V_{th}^+$ = +2.5 V are clearly observed, as shown in Figure 2c. Obviously, both of the *R-V* hysteresis loop and *P-V* loops exhibit almost the same features such as a significant asymmetric characterization and very close value of $V_{th}^+$ ($V_{th}^-$), demonstrating hysteretic modulation of the device resistance maybe correlate to the ferroelectric polarization reversal. To further evaluate the correlation between the polarization reversal and RS behavior, we investigate the *I-V* curves at high frequency, as shown in Figure 2d. In this measurement, the voltage was swept as +5 V → -5 V → +5 V at the frequency of 300 kHz. Positive and negative current peaks occur at *V* = +2.0 V and -4.0 V respectively, at a frequency of 300 kHz, whereas the current peaks become invisible at a reduced frequency of 1 kHz (Figure S2 in the Supporting information), suggesting that the origin of the current peaks is a ferroelectric displacement current.[45, 46] It should be pointed out that the memristor resistance changes from HRS to LRS after the positive current peak, whereas it changes from LRS to HRS after the negative current peak. Considering the mostly initial downward polarization direction of BFO (point to SRO), polarization reversal between downward and upward (point to BFO surface) is expected to be found during the voltage sweep. More importantly, the location of the current peaks is consistent to the coercive voltages in both of $R$–$V$ curve and saturated $P$–$V$ loop. Therefore, we proposed that the RS behavior is attributed to the polarization reversal of BFO layer. To confirm it, we investigate the dependence of local conductivity on ferroelectric polarization direction by combining PFM and conductive Atomic force microscopy (CAFM) measurement in a thinner BFO film. As shown in Figure 2a, we applied a tip bias of -5 V in 2×2 μm$^2$ to switch the polarization upward and then applied +5 V in 1×1 μm$^2$ to switch the polarization downward. The bottom part of Figure 2a is the resulting out-of-plane phase image which shows 180 ° phase difference between the areas poled by +5 V and -5 V due to ferroelectric polarization reversal. After poling, a current map is acquired by scanning the polarization-patterned area with a dc bias of 0.5 V, as shown in the top part of Figure 2a, and it matches well with the PFM domain patterns on the top part. A larger current is observed in the central domain with a downward ferroelectric polarization, which is in consistent with Figure 2b. Based on above experimental results, we confirm that BFO memristor RS mechanism is ambiguously controlled by ferroelectric polarization.

We further analyze the underlying mechanism for ferroelectricity controlled memristor behavior. As can be seen in Figure 1b and S3, BFO memristor exhibits a diode-like rectifying characteristics, indicating formation of Schottky-like barrier. A large value of ideality factors extracted from the *I-V* curves further confirms the formation of an interfacial layer (Figure S4 in Supporting information).[47, 48] Moreover, the rectifying behavior could be modified by switching the BFO polarization direction, as shown in Figure S3. Therefore, we attribute the variation of the rectifying characterization to the potential profile modification of a Schottky-like barrier at the Au/BFO interface due to the polarization reversal. It has been reported that a $V_O$-rich defective layer remains between the Au and BFO layers due to migration of $V_O$ toward the top surface to compensate for the negative polarization charge when BFO exhibits an initial downward self-polarization.[49-51] Whereas, the epitaxial SRO/BFO interface always shows Ohmic contact characteristics.[45] As shown in Figure 2e, the potential profile at the Au/interfacial layer/BFO (Au/IL/BFO) interface significantly depends on the polarization direction of BFO. For the HRS state (downward polarization), large band bending appears in the interfacial layer as well as in the depletion layer at the Au/IL/BFO interface. On the other hand, for the LRS state (upward polarization), the interfacial layer is bent oppositely at the Au/IL/BFO interface, since the opposite polarization charge appears at the interface. The interfacial layer and the depletion layer act as an effective barrier to the hole carrier conduction at the Au/IL/BFO interface and the effective barrier height in the downward polarization state is higher than that in the upward polarization state. The polarization directions dependent barrier height controls hole-carrier conduction at the Au/BFO interface, which is account for the BFO memristor resistive switching effect.

Basing on the above results, it is clear that the memristor behaviors origin from the ferroelectricity of BFO layer. It should be noted that multi-level resistance states have been observed, as shown in Figure 1b. More importantly, the intermediate states exhibit a good cycle to cycle reproducibility up to $10^4$ cycles (Figure S4). Considering the continuous variation of resistance, good fatigue behavior, the BFO memristor is expected to be a promising candidate for implementing nature of synaptic weight.

**Broad Temperature Range Adaptability and High Temperature Ferroelectricity Characteristic**

As shown in Figure 1b, multi-level resistance states have been achieved by varying amplitude of voltage at RT, therefore, BFO memristor is expected to be a promising device to mimic the biological synapse at the temperature where ferroelectricity of BFO would be obtained. Considering the robust ferroelectricity of bulk BFO due to its extremely high Curie temperature of ~1103 K,[42] the ferroelectric polarization dominated RS characterization is expected to be stable at a broad temperature range. Therefore, the transport measurements were performed from -170 to 300 ℃ to examine the temperature adaptability of the BFO memristor. Figure 3a shows the *I–V* loops in semi-logarithmic scale, and a hysteretic characterization has been observed at negative voltage at all the measurement temperatures, suggesting existence of RS behavior at a broad temperature range. Figure 3b shows the corresponding *R-V* hysteretic loops, and a typical nonvolatile RS characterization has been also clearly found. Even though, the ON/OFF ratio shows a decrease with increase temperature due to the enhancement of leakage current at elevated temperatures, at the highest temperature of 300 ℃, an ON/OFF ratio as high as 1000% could be still achieved. So far, few works have been reported on the RS characterization at such high temperature range. Wang *et al.*[52] reported a stable RS behavior in a fully 2D materials heterostructure of graphene/$MoS_{2-x}O_x$/graphene up to 340 ℃. Memristor performance has been also found in $HfO_2$[53] and BFO-based[46] devices at 200 ℃. However, it should be mentioned that synaptic performance has not been identified so far.

As shown in Figure 3b, the switching voltage of transition between HRS and LRS is quite stable as $V_{th}^+$/ $V_{th}^-$ = +2 V/-4 V in all the measurement temperature, which is consistent with the coercive voltage during the convention *P-E* measurement as shown in Figure 2c. Even though the high Curie temperature of bulk BFO, previous reports always focused on the RT, and investigation on ferroelectricity of BFO film at an elevated temperature is still a challenge, due to the high leakage current of BFO compounds.[54, 55] Since it is difficult to obtain ferroelectric hysteresis loops by conventional *P-E* loops at elevated temperature, PFM measurement was carried out to investigate ferroelectricity of BFO thin film in the current work. A poling process (-10 V/4×4 μm$^2$ →+10 V/2×2 μm$^2$→-10 V/1×1 μm$^2$) was firstly performed with a tip bias at RT. Figure 4a shows the following PFM out-plane (OP) phase image measured at RT after poling, after heating at 300 ℃/5 min, and after heating at 500 ℃/5 min, respectively. Even though, a phase contrast of 180 ° could be clearly observed, ferroelectricity of BFO at elevated temperature could not be unambiguously identified.[56-59] For example, Balke and Guan *et al.* reported a ferroelectric-like PFM results in nonferroelectric materials such as $Al_2O_3$ and $HfO_2$.[57, 58] Bark *et al.* reported that PFM signals can be observed due to electrostatic effects caused by the presence of surface charges, or effects related to $V_O$ migration under the presence of tip voltage in $LaAlO_3/SrTiO_3$ hetero-structures.[59] To further confirm the ferroelectricity of BFO films, the retention of PFM signal was investigated, and PFM measurement was carried out after heating at 300 ℃ from 15 to 120 minutes, as shown in Figure 4b, and the corresponding SKPM results also probe the intrisinc ferroelectricity, as shown in Figure S5. Combining the results of Figure 4 and S5, it is found that the PFM phase and SKPM images both show no obvious degradation even after heating for 500 ℃/5 min and 300 ℃/120 min, demonstrating the PFM results origin from an intrinsic robust ferroelectric characterization. The demonstrated stable temperature of 500 ℃ is record-high for ferroelectric polarization of BFO films, and open a door for the applications of ferroelectric thin films at high-temperature harsh electronics.

**Synaptic functions**

Figure 5a shows a schematic of the pulse train used for one measurements, and a pulse train consists of a single reset pulse ($V_{reset}$ = +6/-6 V with pulse duration $t_{reset}$ = 5 ms) to initialize resistance state and 100 identical consecutive pulses ($V_{set}$ = -7～+6 V, pulse duration $t_{set}$ = 0.2～50 μs and pulse interval $t_{interval}$ = 1 ms). A voltage pulse ($V_{read}$ = -1.5 V/0.5 ms) is applied after each programming pulse to read the resistance of BFO memristors. Figure 5b shows the resistance change with various $t_{set}$ as number of programming pulse for $V_{set}$ = +6/-6 V, and Figure 5c shows the change with different $V_{set}$ amplitude as number of programming pulse for $t_{set}$ = 500 ns. It is found that resistance modulation behavior would be well controlled by pulse amplitude and width. Little change has been observed for $V_{set}$ with low amplitude or short $t_{set}$, whereas, an abrupt enhancement of resistance without intermediate resistance state is found for $V_{set}$ with high amplitude or long $t_{set}$. A continuous variation of resistance would be obtained by optimizing amplitude and width of $V_{set}$, and suitable positive/negative pulse could be used to as potentiation/depression spike to change synapse weight. *LTP* and *LTD* behaviors have been clearly seen in Figure 5b, c. By using consecutive *P/D* pulses, a continual tuning of weight states has been realized, and Figure 5d shows the resistance variation after applying the fixed training pulse, including 100 identical potentiation pulses (+6 V/500 ns) and 100 identical depression pulses (-6 V/500 ns) at RT, and a nonlinear *P/D* response due to the training pulses could be clearly found.

Role of programming pulse number on the resistance was then studied. As shown in Figure 5f, a pulse programming that trained with increasing the number of positive pulses and fixed the number of consecutive negative pulses was applied, and the resistance was read after each pulse. It is observed that the device resistance shows a gradually increase with the number of positive pulses, and the final resistance after each negative pulses sequence depends on the level of the previous positive pulse train. Similar results were also found by a pulse training consisting of fixed the number of positive pulses and increasing the number of negative pulses. It means that higher (lower) resistance states could be achieved by fixed number of *D* pulses with increasing number of *P* pulses (fixed number of *P* pulses with increasing number of *D* pulses), as shown in Figures 5f, g. In a word, the resistance modulation could be also achieved by varying number of applied identical pulse, which is more appropriate for the hardware implementation of the brain-inspired computing system.[60] Overall, the conductance can be well controlled by varying the simulation pulse, including width, amplitude and number, which serve as the driving force of domain gradual switching in BFO memristor.

In the above part, synaptic plasticity at RT has been clearly identified. Considering memristive characterization originated from the robust ferroelectric polarization of BFO films which are stable at a broad temperature range, as shown in Figure 4, the *P/D* characterization has been further studied at high temperature. Figure 5e shows that the device resistance can be incrementally adjusted by tuning the duration and sequence of the applied programming voltage, suggesting a well linear synaptic plasticity in the prepared BFO memristor at a temperature as high as 300 ℃.

As known, one important biologically plausible learning rule to update the synaptic weight is the *STDP* characterization, which states that if the presynaptic neuron spikes earlier than the postsynaptic neuron ($\Delta t > 0$), the weight of the synapse increase and vice versa.[61] The smaller timing difference between the two neurons spiking, the larger weight variation is observed. In order to simulate the *STDP* characterization, special waveforms were designed to emulate the pre-and post-synaptic neuron spikes, and Figure 6 insert shows the superposition produces when they reach the memrisotr with a timing difference of Δ*t*. As can be seen from the experimental *STDP* curve, only small |Δ*t*| produce a conductance change whereas large |Δ*t*| leave the device unchanged. Figure 6 shows the *STDP* curves at -170 ℃, 25 ℃ and 300 ℃, respectively. The simulation of the *STDP* is generated based on the Hebbian learning rule as follows:

$$\Delta w = \begin{cases} A_+ \cdot \exp\left(\frac{\Delta t}{\tau}\right), & \Delta t > 0 \\ A_- \cdot \exp\left(\frac{-\Delta t}{\tau}\right), & \Delta t < 0 \end{cases} \quad (1)$$

$$\Delta w = \frac{G - G_0}{G_0} \times 100\%, \quad \Delta t = t_{pre} - t_{post} \quad (2)$$

Where *A* determines the spike amplitude, $\tau$ is the decay time constant, Δ*t* is represent the spike time interval between the pre and postsynaptic, and the synaptic weight change (Δ*w*) is defined in the equation (2), the conductance is measure before ($G_0$) and after (*G*) the spike-pair application. It could be clearly found that all the experimental data can be well fitted to a biological synapse model , the range of *ΔG* for potentiation is (0, +∞), and for depression, it is (−1, 0).[61] The change in the synaptic weight with the function of Δ*t* could be well-fitted with the exponential functions (1), suggesting their good synaptic *STDP* characteristics in all the measurement temperature range. It should be noted that the effective timing window is about 300 μs, 60 μs and 4 μs for the operation temperature of -170 ℃, 25 ℃ and 300 ℃, respectively. The decrease of the effective timing window with temperature is attributed to the high reversal speed of ferroelectric domain at evaluated temperature. Our current work demonstrates BFO memristor could be a promising emulator in a temperature up to 300 ℃, and we expected that a much higher operation temperature even up to 500 ℃ if the leakage current would be further suppressed by doping,[32, 54] interface engineering,[62] or decreasing the device size [63] to achieve synaptic functions at much higher temperature.

**Conclusion**

A robust ferroelectricity has been unambiguously identified at an elevated temperature as high as 500 ℃ in the high quality epitaxial BiFeO$_3$ (BFO) thin films. We have realized various synaptic learning and memory functions including *LTP*, *LTD*, *P/D*, and *STDP* based on the prepared BFO memristor in a wide range of temperature from -170 to 300 ℃. Not only the identical hysteresis loop of the *R-V* and *P-V* characterization both also the consistent voltage for ferroelectric displacement current peaks and RS threshold voltages is observed in the BFO memristor. It provides strong evidence that the RS behavior was attributed to the barrier profile modification at the Au/BFO interface due to ferroelectric polarization reversal of BFO layer. The memristor based on robust ferroelectricity with an extremely thermal stability provides a feasible way for the realization of reliable synaptic devices in extreme wide temperature applications.

**Experimental Section**

*Film Fabrication:* Epitaxial BFO thin films were fabricated on (001)-oriented SrTiO$_3$ (STO) substrates with ~ 40 nm-thick SRO buffer layers by pulsed laser deposition using a KrF excimer laser ($\lambda$ = 248 nm). The SRO films were firstly deposited at 650 ℃ under an oxygen background pressure of 13 Pa, following by deposition of the ~70 nm-thick BFO films. The $Bi_{1.1}FeO_3$ target was ablated at a fluence of 0.9 J/cm$^2$ and a repetition rate of 10 Hz. And the metallic SRO films not only sever as bottom electrodes, but also provide step-and-terrace surfaces for growing atomically flat BFO films. The films were then cooled down in oxygen atmosphere of $5\times10^4$ Pa at a cooling rate of 20 ℃/min. Au (100 nm) with a square of $25 \times 25$ μm$^2$ were evaporated on the BFO film as top electrode through thermal evaporation and lithography process.

*Structural Analysis:* An atomic force microscope (AFM, Veeco Dimension 3100) was used to characterize the surface morphology of the films. The high-resolution X-ray diffraction (XRD, Bruker D8 Discover) pattern of epitaxial thin films was performed to evaluate the lattice parameters.

*Device Characterization:* All electric measurements of BFO memristor were performed in a home-made probing station from -170 to 300 ℃. Ferroelectric hysteresis loop was carried out by Precision LC ferroelectric tester (Radiant Technology) at -170 ℃. Electrical and pulse measurements were measured with a Keithley Semiconductor Parameter Analyzer (4200A-SCS) with pulse measuring units. Room-temperature PFM measurement was conducted using a commercial scanning probe microscope (SPM) system (Asylum Research Cypher) with a software platform (IGOR PRO 6.12A) and SPM control software. A commercially available Cr-Au coated Si tip (NSC36, MikroMasch) was employed for all the measurement. PFM switching spectroscopy (PFS) was performed to analyze the local switching behavior of the BFO thin films.

**Acknowledgements**

This work was supported by the National Key Research and Development Program of China (Grant No. 2017YFA0303403), the National Natural Science Foundation of China (Grant Nos. 11304097，11874149, 11774092 and 51572085), Shanghai Science and Technology Innovation Action Plan (Grant No. 17JC1402500), and the Fundamental Research Funds for the Central Universities.
**Acknowledgements**

This work was supported by the National Key Research and Development Program of China (Grant No. 2017YFA0303403), the National Natural Science Foundation of China (Grant Nos. 11304097，11874149, 11774092 and 51572085), Shanghai Science and Technology Innovation Action Plan (Grant No. 17JC1402500), and the Fundamental Research Funds for the Central Universities.

References

[1]    P. A. Merolla, J. V. Arthur, R. Alvarez-Icaza, A. S. Cassidy, J. Sawada, F. Akopyan, B. L. Jackson, N. Imam, C. Guo, Y. Nakamura, B. Brezzo, I. Vo, S. K. Esser, R. Appuswamy, B. Taba, A. Amir, M. D. Flickner, W. P. Risk, R. Manohar, D. S. Modha, *Science* **2014**, 345, 668.

[2]    K. Amunts, C. Lepage, L. Borgeat, H. Mohlberg, T. Dickscheid, M. E. Rousseau, S. Bludau, P. L. Bazin, L. B. Lewis, A. M. Oros-Peusquens, N. J. Shah, T. Lippert, K. Zilles, A. C. Evans, *Science* **2013**, 340, 1472.

[3]    D. C. Van Essen, *Neuron* **2013**, 80, 775.

[4]    D. Silver, A. Huang, C. J. Maddison, A. Guez, L. Sifre, G. van den Driessche, J. Schrittwieser, I. Antonoglou, V. Panneershelvam, M. Lanctot, S. Dieleman, D. Grewe, J. Nham, N. Kalchbrenner, I. Sutskever, T. Lillicrap, M. Leach, K. Kavukcuoglu, T. Graepel, D. Hassabis, *Nature* **2016**, 529, 484.

[5]    D. A. Drachman, *Neurology* **2005**, 64, 2004.

[6]     E. R. Kandel, *Science* **2001**, 294, 1030.

[7]     G. Voglis, N. Tavernarakis, *EMBO Rep.* **2006**, 7, 1104.

[8]     G.-q. Bi, M.-m. Poo, *Nature* **1999**, 401, 792.

[9]     J. J. Yang, D. B. Strukov, D. R. Stewart, *Nat. Nanotechnol.* **2013**, 8, 13.

[10]    Y. LeCun, Y. Bengio, G. Hinton, *Nature* **2015**, 521, 436.

[11]    M. Prezioso, F. Merrikh-Bayat, B. D. Hoskins, G. C. Adam, K. K. Likharev, D. B. Strukov, *Nature* **2015**, 521, 61.

[12]    Z. Wang, S. Joshi, S. Savel'ev, W. Song, R. Midya, Y. Li, M. Rao, P. Yan, S. Asapu, Y. Zhuo, *Nat. Electron.* **2018**, 1, 137.

[13]    M. A. Zidan, J. P. Strachan, W. D. Lu, *Nat. Electron.* **2018**, 1, 22.

[14]    G. Indiveri, S.-C. Liu, *Proc. IEEE* **2015**, 103, 1379.

[15]    S. Yu, *Proc. IEEE* **2018**, 106, 260.

[16]    P. Stoliar, J. Tranchant, B. Corraze, E. Janod, M.-P. Besland, F. Tesler, M. Rozenberg, L. Cario, *Adv. Funct. Mater.* **2017**, 27, 1604740.

[17]    F. Tesler, C. Adda, J. Tranchant, B. Corraze, E. Janod, L. Cario, P. Stoliar, M. Rozenberg, *Phys. Rev. Appl.* **2018**, 10, 054001.

[18]    S. Kumar, J. P. Strachan, R. S. Williams, *Nature* **2017**, 548, 318.

[19]    S. Ambrogio, N. Ciocchini, M. Laudato, V. Milo, A. Pirovano, P. Fantini, D. Ielmini, *Front. Neurosci.* **2016**, 10, 56.

[20]    S. B. Eryilmaz, D. Kuzum, R. Jeyasingh, S. Kim, M. BrightSky, C. Lam, H. S. Wong, *Front. Neurosci.* **2014**, 8, 205.

[21]    A. Sebastian, T. Tuma, N. Papandreou, M. Le Gallo, L. Kull, T. Parnell, E.

Eleftheriou, *Nat. Commun.* **2017**, 8, 1115.

[22] M. Hu, C. E. Graves, C. Li, Y. Li, N. Ge, E. Montgomery, N. Davila, H. Jiang, R. S. Williams, J. J. Yang, Q. Xia, J. P. Strachan, *Adv. Mater.* **2018**, 30, 1705914.

[23] M. M. Shulaker, G. Hills, R. S. Park, R. T. Howe, K. Saraswat, H. P. Wong, S. Mitra, *Nature* **2017**, 547, 74.

[24] A. Serb, J. Bill, A. Khiat, R. Berdan, R. Legenstein, T. Prodromakis, *Nat. Commun.* **2016**, 7, 12611.

[25] F. Miao, J. P. Strachan, J. J. Yang, M. X. Zhang, I. Goldfarb, A. C. Torrezan, P. Eschbach, R. D. Kelley, G. Medeiros-Ribeiro, R. S. Williams, *Adv. Mater.* **2011**, 23, 5633.

[26] D. H. Kwon, K. M. Kim, J. H. Jang, J. M. Jeon, M. H. Lee, G. H. Kim, X. S. Li, G. S. Park, B. Lee, S. Han, M. Kim, C. S. Hwang, *Nat. Nanotechnol.* **2010**, 5, 148.

[27] Q. Liu, J. Sun, H. Lv, S. Long, K. Yin, N. Wan, Y. Li, L. Sun, M. Liu, *Adv. Mater.* **2012**, 24, 1844.

[28] C. Chen, C. Song, J. Yang, F. Zeng, F. Pan, *Appl. Phys. Lett.* **2012**, 100, 253509.

[29] R. E. Simpson, M. Krbal, P. Fons, A. V. Kolobov, J. Tominaga, T. Uruga, H. Tanida, *Nano Lett.* **2010**, 10, 414.

[30] K. Ohara, L. Temleitner, K. Sugimoto, S. Kohara, T. Matsunaga, L. Pusztai, M. Itou, H. Ohsumi, R. Kojima, N. Yamada, *Adv. Funct. Mater.* **2012**, 22, 2251.

[31]    H.-S. P. Wong, S. Raoux, S. Kim, J. Liang, J. P. Reifenberg, B. Rajendran, M. Asheghi, K. E. Goodson, *Proc. IEEE* **2010**, 98, 2201.

[32]    Y.-F. Tan, Y.-T. Su, M.-C. Chen, T.-C. Chang, T.-M. Tsai, Y.-T. Tseng, C.-C. Yang, H.-X. Zheng, W.-C. Chen, C.-C. Lin, X.-H. Ma, Y. Hao, S. M. Sze, *Appl. Phys. Express* **2019**, 12, 024004.

[33]    C. Lu, J. Yu, X.-W. Chi, G.-Y. Lin, X.-L. Lan, W. Huang, J.-Y. Wang, J.-F. Xu, C. Wang, C. Li, S.-Y. Chen, C. Liu, H.-K. Lai, *Appl. Phys. Express* **2016**, 9, 041501.

[34]    A. Chanthbouala, V. Garcia, R. O. Cherifi, K. Bouzehouane, S. Fusil, X. Moya, S. Xavier, H. Yamada, C. Deranlot, N. D. Mathur, M. Bibes, A. Barthelemy, J. Grollier, *Nat. Mater.* **2012**, 11, 860.

[35]    D. J. Kim, H. Lu, S. Ryu, C. W. Bark, C. B. Eom, E. Y. Tsymbal, A. Gruverman, *Nano Lett.* **2012**, 12, 5697.

[36]    V. Garcia, M. Bibes, *Nat. Commun.* **2014**, 5, 4289.

[37]    B. Tian, L. Liu, M. Yan, J. Wang, Q. Zhao, N. Zhong, P. Xiang, L. Sun, H. Peng, H. Shen, *Adv. Electron. Mater.* **2018**, 5, 1800600.

[38]    L. W. Martin, A. M. Rappe, *Nat. Rev. Mater.* **2017**, 2, 16087.

[39]    M. Dawber, K. Rabe, J. Scott, *Rev. Mod. Phys.* **2005**, 77, 1083.

[40]    H. Yamada, V. Garcia, S. p. Fusil, S. r. Boyn, M. Marinova, A. Gloter, S. p. Xavier, J. Grollier, E. Jacquet, C. c. Carrétéro, *ACS nano* **2013**, 7, 5385.

[41]    S. Boyn, J. Grollier, G. Lecerf, B. Xu, N. Locatelli, S. Fusil, S. Girod, C. Carretero, K. Garcia, S. Xavier, J. Tomas, L. Bellaiche, M. Bibes, A.

Barthelemy, S. Saighi, V. Garcia, *Nat. Commun.* **2017**, 8, 14736.

[42] J. Wang, J. Neaton, H. Zheng, V. Nagarajan, S. Ogale, B. Liu, D. Viehland, V. Vaithyanathan, D. Schlom, U. Waghmare, *Science* **2003**, 299, 1719.

[43] D. Kuzum, S. Yu, H. S. Wong, *Nanotechnology* **2013**, 24, 382001.

[44] R. S. Zucker, W. G. Regehr, *Annu. Rev. Phys.* **2002**, 64, 355.

[45] A. Tsurumaki, H. Yamada, A. Sawa, *Adv. Funct. Mater.* **2012**, 22, 1040.

[46] A. Tsurumaki-Fukuchi, H. Yamada, A. Sawa, *Appl. Phys. Lett.* **2013**, 103, 152903.

[47] T. Choi, S. Lee, Y. Choi, V. Kiryukhin, S.-W. Cheong, *Science* **2009**, 324, 63.

[48] M. Ben‐Chorin, F. Möller, F. Koch, *J. Appl. Phys.* **1995**, 77, 4482.

[49] D. Lee, S. H. Baek, T. H. Kim, J. G. Yoon, C. M. Folkman, C. B. Eom, T. W. Noh, *Physical Review B* **2011**, 84, 125305.

[50] M. F. Chisholm, W. Luo, M. P. Oxley, S. T. Pantelides, H. N. Lee, *Phys. Rev. Lett.* **2010**, 105, 197602.

[51] Y. M. Kim, A. Morozovska, E. Eliseev, M. P. Oxley, R. Mishra, S. M. Selbach, T. Grande, S. T. Pantelides, S. V. Kalinin, A. Y. Borisevich, *Nat. Mater.* **2014**, 13, 1019.

[52] M. Wang, S. Cai, C. Pan, C. Wang, X. Lian, Y. Zhuo, K. Xu, T. Cao, X. Pan, B. Wang, S.-J. Liang, J. J. Yang, P. Wang, F. Miao, *Nat. Electron.* **2018**, 1, 130.

[53] H. Lee, P. Chen, T. Wu, Y. Chen, C. Wang, P. Tzeng, C. Lin, F. Chen, C. Lien, M.-J. Tsai, presented at 2008 IEEE International Electron Devices Meeting **2008**.

[54] D. Lebeugle, D. Colson, A. Forget, M. Viret, P. Bonville, J. F. Marucco, S. Fusil, *Phys. Rev. B* **2007**, 76, 024116.

[55] J. Lv, X. Lou, J. Wu, *J. Mater. Chem. C* **2016**, 4, 6140.

[56] D. Denning, J. Guyonnet, B. J. Rodriguez, *Int. Mater. Rev.* **2016**, 61, 46.

[57] Z. Guan, Z.-Z. Jiang, B.-B. Tian, Y.-P. Zhu, P.-H. Xiang, N. Zhong, C.-G. Duan, J.-H. Chu, *AIP Adv.* **2017**, 7, 095116.

[58] N. Balke, P. Maksymovych, S. Jesse, A. Herklotz, A. Tselev, C.-B. Eom, I. I. Kravchenko, P. Yu, S. V. Kalinin, *ACS Nano* **2015**, 9, 6484.

[59] C. W. Bark, P. Sharma, Y. Wang, S. H. Baek, S. Lee, S. Ryu, C. M. Folkman, T. R. Paudel, A. Kumar, S. V. Kalinin, A. Sokolov, E. Y. Tsymbal, M. S. Rzchowski, A. Gruverman, C. B. Eom, *Nano Lett.* **2012**, 12, 1765.

[60] S. Yu, *Neuro-inspired computing using resistive synaptic devices*, Springer, **2017**.

[61] G.-q. Bi, M.-m. Poo, *J. Neurosci.* **1998**, 18, 10464.

[62] W. Chen, N. Chamele, Y. Gonzalez-Velo, H. J. Barnaby, M. N. Kozicki, *IEEE Electron Device Lett.* **2017**, 38, 1244.

[63] A. R. Damodaran, E. Breckenfeld, Z. Chen, S. Lee, L. W. Martin, *Adv. Mater.* **2014**, 26, 6341.

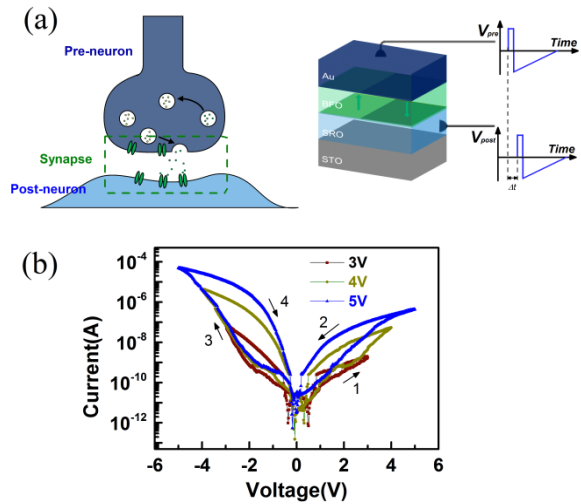

Figure 1. Characterization and electrical characteristics of the BFO memristor. a) Schematic illustration of the concept of using ferroelectric memristor as synapses between neurons. The ferroelectric memristor of $BiFeO_3$ (BFO) is sandwiched between bottom electrode of $SrRuO_3$ (SRO) and top electrode of Au. STO stands for $SrTiO_3$. The synaptic transmission is modulated by the causality ($\Delta t$) of neuron spikes. The synaptic weight (channel conductance) in BFO memristor is controlled by the average ferroelectric polarization. b) *I-V* curves measured on BFO memristor by increasing the voltage sweep range. The numbers denote the sequence of voltage sweeps.

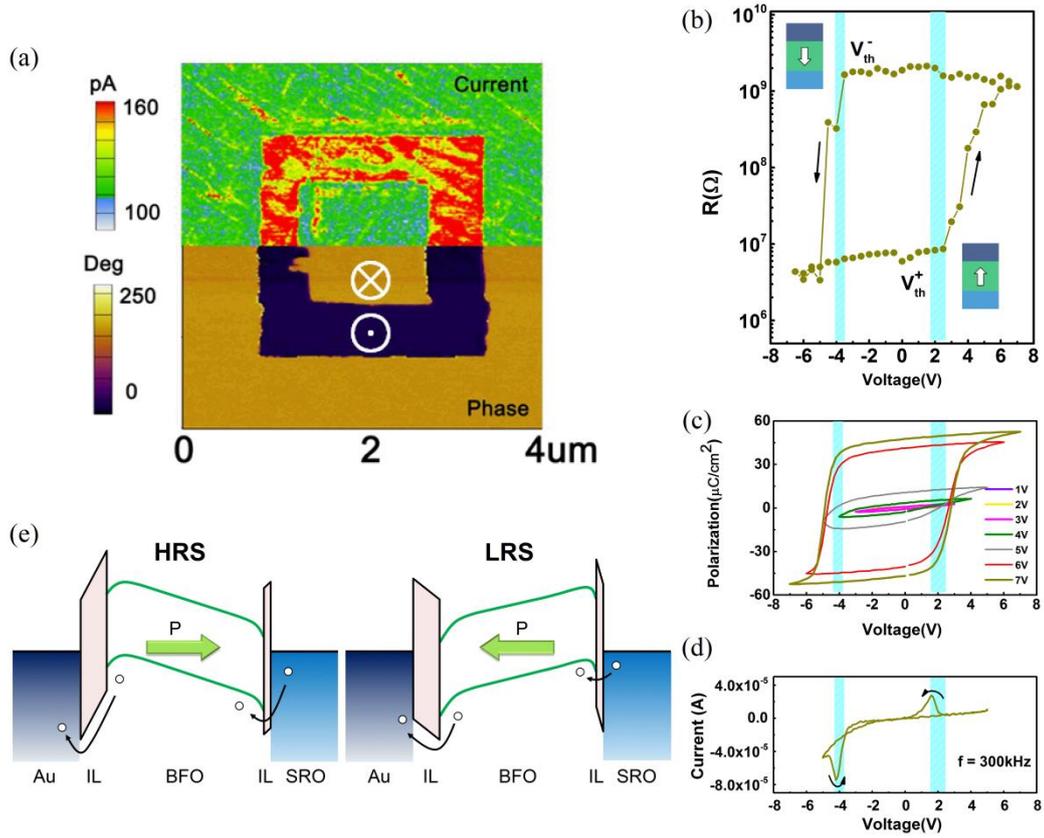

Figure 2. The correlation between polarization reversal and resistance switching of BFO memristor. a) The lower half part is out-of-plane phase image acquired after poling sequence of -5 V / 2×2 μm$^2$ and +5 V / 1×1 μm$^2$ and the upper half part is the corresponding current mapping measured at -0.5 V. b) Resistance versus pulse voltage hysteresis loop of the BFO memristor displaying clear voltage thresholds ($V_{th}^+$ and $V_{th}^-$). c) P–V hysteresis loops with different maximum voltages of the BFO memristor at 100 K. d) I–V characteristics of BFO memristor measured at 300 kHz. e) Schematic energy-band diagrams of Au/BFO/SRO memristor with interfacial layers (ILs) at V= 0. In the HRS, a downward polarization state increases the band bending of the interfacial layer and the depletion layer with raising the effective barrier height for hole carrier conduction. In the LRS, an upward polarization state decreases the effective barrier height.

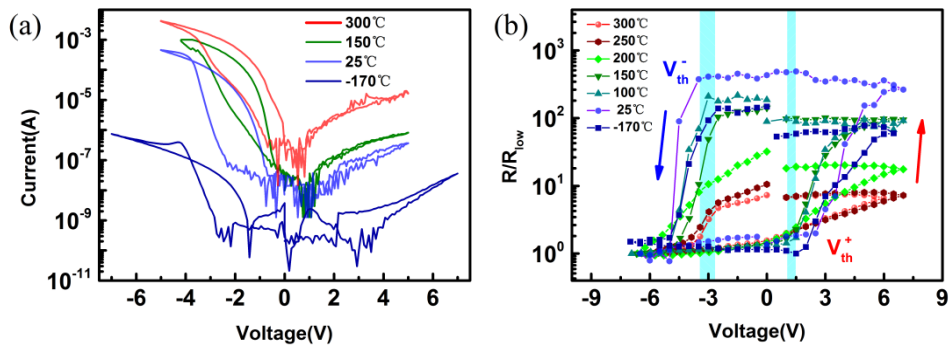

Figure 3. Temperature dependence of electrical characteristics of BFO memristor. a) Pulse *I–V* curves sweep in a wide range of temperature from -170 to 300 ℃. b) Temperature-dependent of *R-V* hysteresis loops displaying identical voltage thresholds ($V_{th}^{+}$ and $V_{th}^{-}$) under different temperature.

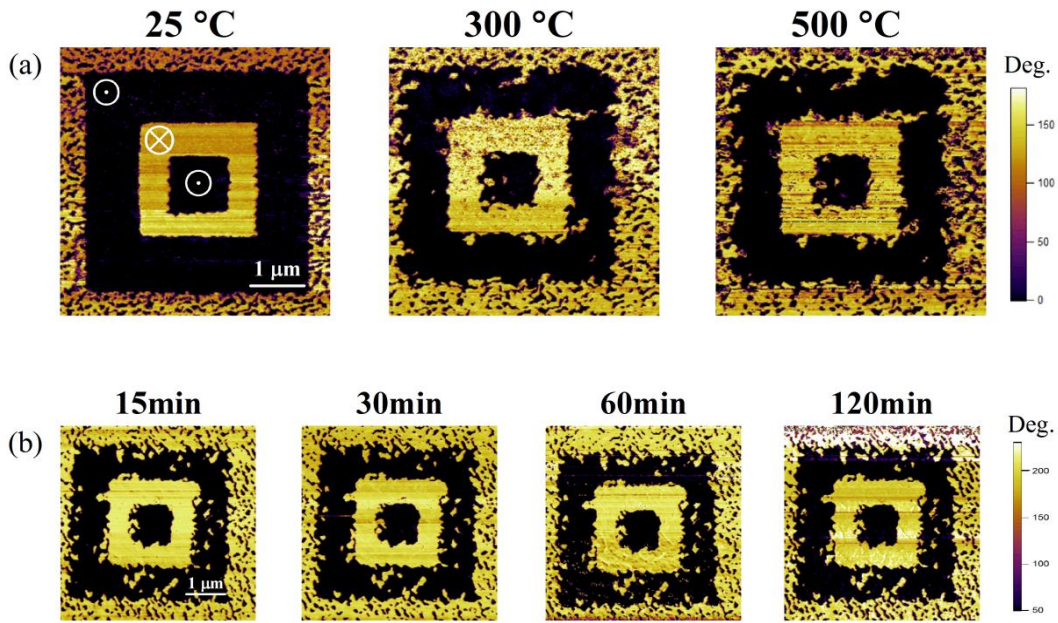

Figure 4. Ferroelectric characterization of BFO films at elevated temperature by PFM measurement. a) out-of-phase images acquired after poling sequence of -10 V/4×4 μm$^2$ and 1×1 μm$^2$ and +10 V/2×2 μm$^2$ at RT. The phase images measured at RT after heating at 25, 300 and 500 ℃ for 5 min, respectively. b) PFM phase images after the poling process heated at 300 ℃ for different time intervals: 15 min, 30 min, 60 min and 120 min, respectively.

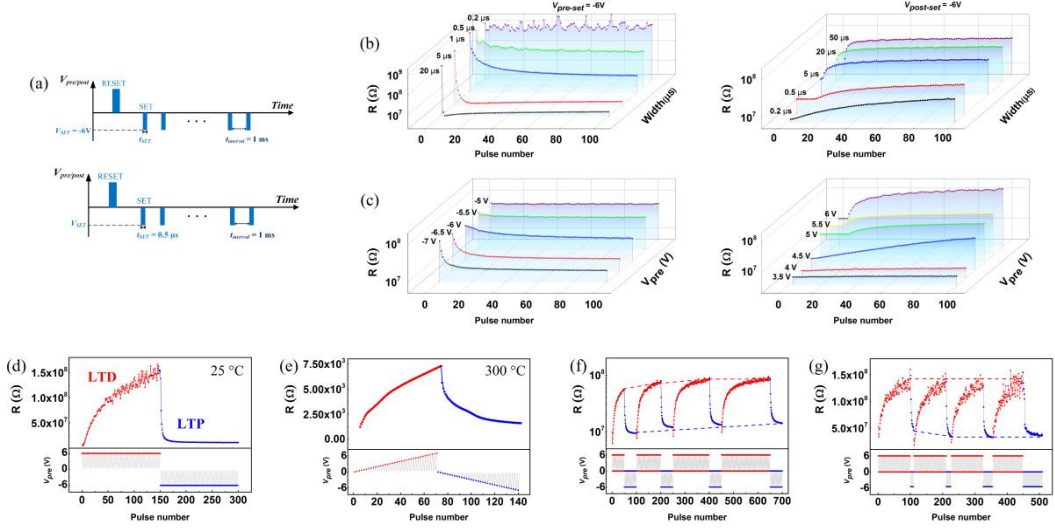

Figure 5. Synaptic functions of *LTP/LTD* in BFO memristor. a) Schematic of pulse trains used for the measurements. A pulse train consists of a single reset pulse ($V_{pre/post-reset}$ = 6 V, pulse duration $t_{reset}$ = 5 ms) and 100 subsequent set pulses with different b) pulse widths ($V_{pre/post-set}$ = -6 V) or c) voltage amplitudes (pulse width $t_{set}$ = 500 ns). The platform of conductance is due to reach the current range limit. Emulation of synaptic plasticity with BFO memristor. d) The evolution of the resistance realized by applying trains of identical depression pulses of +6 V followed by identical potentiation pulses of −6 V (pulse width 500 ns) at 25 ℃. e) The evolution of the resistance by applying trains of increasing/decreasing voltage amplitude pulses with a fixed time width of 500 ns at 300 ℃. Significantly improved synaptic linearity based on pulse engineering. f) and g) Pulse number-dependent resistance change is achieved by using various numbers of training pulses. Higher and lower resistance states can be achieved by increasing the *P* and *D* pulse numbers, respectively. Each write pulse is followed by a read pulse (-1.5 V, 0.5 ms).

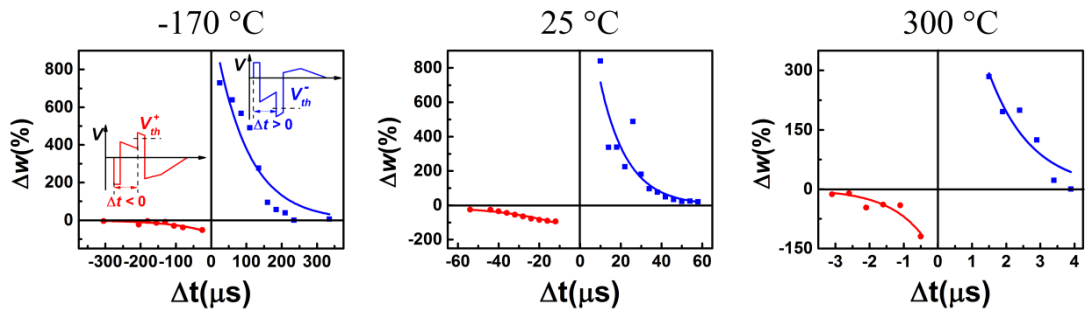

Figure 6. *STDP* properties of three different temperatures (-170, 25 and 300 ℃) in BFO memristor. Modulation of the weight (Δ*w*) as a function of the delay (Δ*t*) between pre- and postsynaptic spikes. The solid lines are the exponential fits to the experimental data. With the temperature increasing, the delay (Δ*t*) of BFO memristor decreased remarkably. Inset shows the pairs of the pre- and post-synaptic spikes, which are designed to implement *STDP*.